\documentclass[12pt,preprint]{aastex}
\def\deg{$^{\circ}~$}
\newcommand{\tnm}{\tablenotemark}
\newcommand{\tnt}{\tablenotetext}
\newcommand{\ks}{$K_{\rm s}$}

\newcommand{\etal}{et~al.}

\newcommand{\mdot}{$\dot{M}$}

\newcommand{\mum}{$\mu$m}

\shorttitle{Spitzer Observations of Eta Cha Association}
\shortauthors{T. N. Gautier et al.}

\begin{document}

\title{Spitzer--MIPS Observations of the $\eta$ Cha Young Association}


\author{Thomas. N. Gautier III\altaffilmark{1,2},
L.M. Rebull\altaffilmark{3},
K.R. Stapelfeldt\altaffilmark{1} and 
A. Mainzer\altaffilmark{1}}

\altaffiltext{1}{Jet Propulsion Laboratory, California Institute of 
Technology, 4800 Oak Grove Drive, Pasadena, CA 91109 USA}
\altaffiltext{2}{email: tngautier@jpl.nasa.gov}
\altaffiltext{3}{Spitzer Science Center, 1200 E. California Blvd,
Pasadena, CA 91125 USA}

\begin{abstract}

We have mapped the $\eta$ Chamaeleontis young stellar association 
in the far-infrared with the Multiband Imaging Photometer for Spitzer 
(MIPS) on the Spitzer Space Telescope.  All sixteen members
within the map region were detected at 24 \mum, along with five members
at 70 \mum\ and two at 160 \mum.  Ten stars show far-infrared excess
emission indicating the presence of circumstellar disks; six of these
have central clearings as evidenced by the onset of excess emission
at $\lambda>$ 5 \mum.  No new infrared excess sources are identified 
among the 113 2MASS field stars with 24 \mum\ photometry but not seen as 
X-ray sources, indicating
that membership lists derived from X-ray surveys are reasonably complete.
Circumstellar disks in the $\eta$ Cha association span the range from 
10$^{-1}$ to 10$^{-4}$ in their fractional infrared luminosity, with a median
L$_d$/L$_*$ of 0.04.  The presence of optically thick, optically thin,
and intermediate optical depth disks within the same stellar group, in
combination with the large fraction of disks with inner holes, indicates 
that the $\eta$ Cha association represents a crucial stage in circumstellar 
disk evolution.

\end{abstract}


\section{Introduction}
\label{intro}

The young stellar association around the B8V star $\eta$
Chamaeleontis  (HD 75416; IRAS F08430-7846) presents an  excellent
opportunity to study the early evolution of circumstellar disks which
may form planetary systems.  The association was discovered by 
\citet{mama99}, and lies at a distance of 97 pc.  Age estimates from 
comparison with stellar evolution models range from 4 to 15 Myr with 
more recent values averaging about 8 Myr (\cite{mama99,ls04, lyo04}). 
\citet{ls04} lists 18 association members, including a late-age
classical  T Tauri star \citep{law02}.  Other members show weak
H$\alpha$ emission,  some with evidence for continuing accretion
\citep{law04}.  A deficit of  wide binaries was noted in the
association by \cite{bran06}.  Three members  (including $\eta$ Cha
itself) have 25 \mum\ excess indicated by the IRAS  Faint Source
Survey (Moshir 1992); two of these were found to have $L$ band  excess
by \cite{hach05}.  The $\eta$ Cha association has already been the 
target of two studies with the {\it Spitzer Space Telescope}. 
\citet{meg05}  targeted 17 members with the Infrared Arrary Camera
(IRAC), and found six  members with 8 \mum\ excess.  \cite{bou06} used
the Infrared Spectrograph  (IRS) to measure the 15 late type members 
from 8-33 \mum, reporting excess in 8 objects. 

In order to probe the outer regions of these disks beyond the ``snow line'' 
where giant planets might potentially form, observations at longer infrared 
wavelengths are needed.  Far-infrared observations are also crucial to inform 
comparisons between disk properties in $\eta$ Cha and the debris disks of
older field stars (which are primarily manifest at $\lambda\ge$ 60 \mum.)
Spitzer's far infrared camera MIPS (Multiband Imaging Photometer for Spitzer;
\cite{rieke04}) provides unprecedented sensitivity in this wavelength region.  
In this contribution, we report the MIPS results for $\eta$ Cha.  

\section{Observations}
\label{obs}

\begin{figure}
\epsscale{1.0}
\plotone{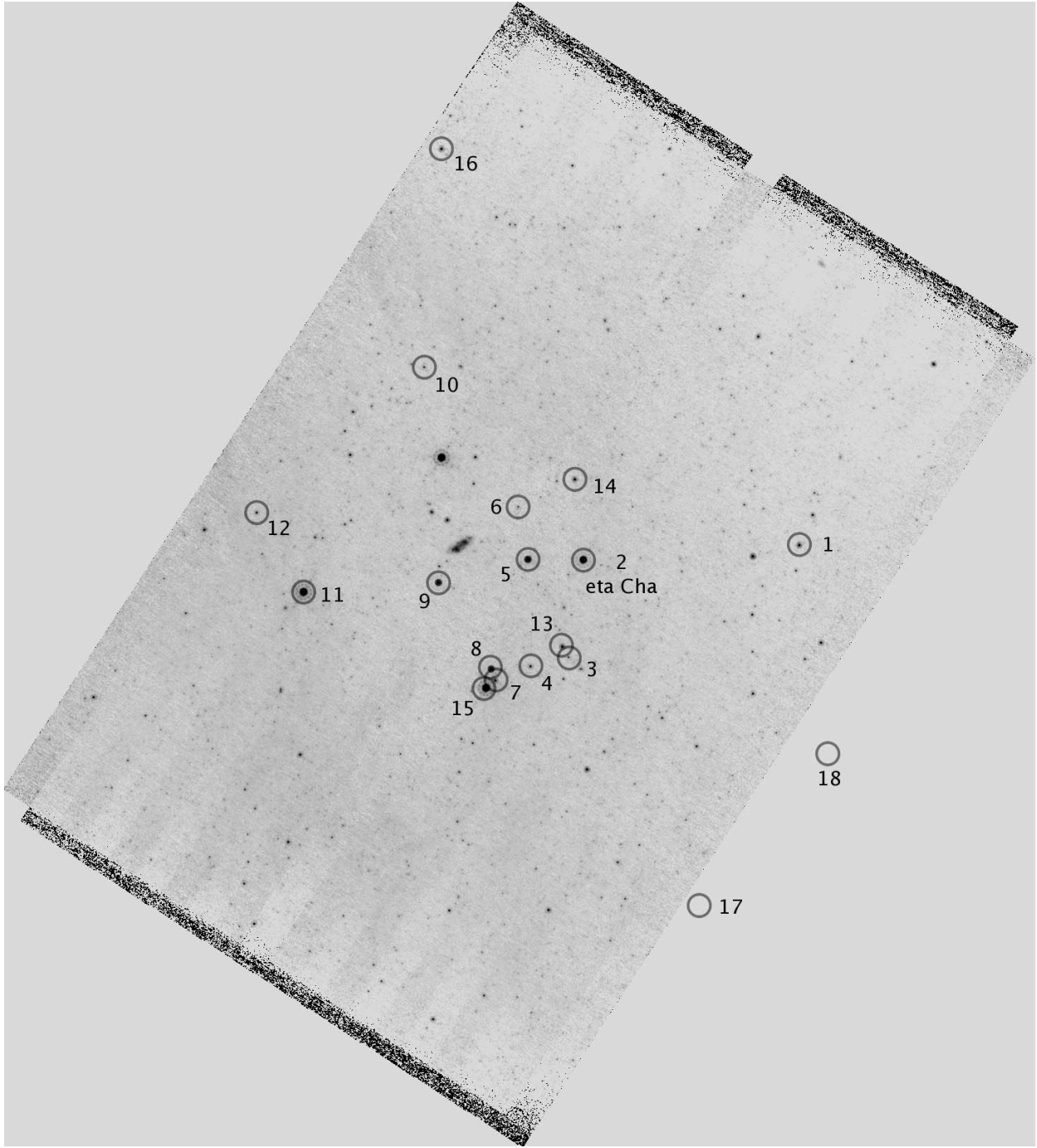}
\caption{
MIPS 24 \mum\ map of the $\eta$ Cha association with various
sources marked. The scan width (upper left to lower right) is
approximately 0.5\deg. North is up, east to the left. 
\label{map_fig}}
\end{figure}

Spitzer/MIPS was was used to map  a $0.5\arcdeg\times0.5\arcdeg$
region covering most of the known members  of the $\eta$ Cha
association.  A ``medium scan'' map with half-array  cross-scan
offsets provided total integration times of 80 sec, 40 sec, and  8 sec
throughout the mapped area at 24, 70, and 160 \mum, respectively.   The
observations were carried out on 2005 April 08 (2.6 hr duration starting at 
JD2453469.11412) under Spitzer program \#100, Spitzer AORKEY 4938752. 
Our complete 24 \mum\ map appears in Figure~\ref{map_fig}.
\clearpage
\clearpage

Table~\ref{tab:data} presents photometry of the $\eta$ Cha association
members as well as upper limits for undetected sources. Note that the
ECHA source numbers correspond to the source ID numbers given in 
\citet{ls04}. ROSAT Eta Cha X-ray (RECX) numbers also correspond for 
objects 1-12. The objects known by ECHA numbers 17 and 18 were not 
measured as they are off the edge of our map.

The 24 \mum\ data were reduced starting from the standard Spitzer
Science Center (SSC) pipeline-produced basic calibrated data (BCD), 
version S13.2. \citep[See][for a description of the pipeline.]{gor05} 
The BCDs suffered from low-level cosmetic defects which were removed
by self-flattening the data as described in the MIPS Data Handbook. 
The MOPEX software package \citep{mak05} was used to re-mosaic the BCDs
at a pixel scale of 2.5\arcsec\ per pixel, very close to the native
pixel scale, and to obtain PSF-fitted source extractions.  Seven hundred fifteen 
sources with SNR $>3$ were extracted from the 24 \mum\ scan map. 
Of the 715 sources, 129 have 2MASS counterparts. All 16 known 
association members that fell within our map were 
detected.  The systematic uncertainty in the 24 \mum\ zero-point 
is estimated to be 4\% \citep{chad07}. The formal $1\sigma$ statistical 
uncertainties are reported in Table~\ref{tab:data}.  An additional assessment 
of our measurement uncertainty appears in \S\ref{kkm24sec}.

For 70 \mum, we started with the filtered BCDs for which
an automated attempt has been made to remove instrumental
signatures.  MOPEX was used to mosaic the BCDs at a pixel scale
of 4\arcsec, about half of the native scale, and to do PSF-fitted
source extraction.  Here again, the $1\sigma$ statistical uncertainties are reported in
Table~\ref{tab:data}. The estimated systematic uncertainty is 10\%
\citep{gor07}.  About 40 sources with SNR $>3$ were extracted at 70 \mum, 
21 with 2MASS counterparts. Five of the known association members were 
detected; upper limits for the remaining observed association members 
were obtained via an examination of the scatter in the background at the 
expected location of the source \citep[following][]{bpmgp}. Those 
3-$\sigma$ upper limits appear in Table~\ref{tab:data}.  ECHA 16, which 
was observed in the 24 \mum\ map, was just off the edge of the 70 \mum\ 
map, so no measurement was obtained for this object.

For 160 \mum\ we followed \citet{bpmgp} starting with the raw BCDs, 
but using the MIPS Data Analysis Tool (DAT) software (version 3.06;
Gordon \etal\ 2005) for the final data reduction.  At 160 \mum\ the map 
was measured only at the locations of known association members. Two 
members were detected. The MIPS 160 \mum\
array suffers from a spectral leak that allows near-IR radiation to
produce a ghost image adjacent to the true 160 micron source image for
stellar temperature, roughly Rayleigh-Jeans, sources brighter than
$J\sim5.5$.  None of the $\eta$ Cha members are as bright as this
limit, so no correction for leak images was needed.  Three-$\sigma$ 
upper limits for undetected members were obtained as in \citet{bpmgp} 
and are given in Table~\ref{tab:data}.

We compiled optical and NIR photometry for the known members
from the literature, primarily 2MASS \citep{2mass} and
\citet{ls04} for shorter wavelengths. For longer wavelengths we used 
the IRAC photometry as reported by \citet{meg05} for the known members 
and IRS photometry reported by \citet{bou06} at 13 and 33 \mum.
For the known cluster members, spectral types are available in
the literature.  These spectral types were used to select a stellar 
photosphere model spectrum from the closest corresponding Kurucz-Lejeune 
model \citep{lej97}, which was then normalized to each object's flux 
density at \ks, except for ECHA 11 which was normalized at J (see 
section \ref{irxs}).  No further manipulation of the spectral fit was made. 
In particular no redding corrections were required, consistent with the 
results of \citet{ls04} and \citet{lyo04}. The normalized photosphere models 
were used to predict the photospheric flux density at 24, 70, and 160 \mum\ 
that are presented in Table~\ref{tab:data}.

In order to band-merge across wavelengths from optical to 160 \mum, we
matched the central positions derived from photometry of each source 
to the expected position of each known association member.  Since the
source density in this field is not high, spurious source matches are
relatively unlikely.  IRAC and 2MASS positions were matched within 1
arcsecond to the MIPS-24 position, and within 2 arcseconds to the
MIPS-70 position.  Photometry for 160 \mum\ was done by hand at the
position of the association member.

Note that the MIPS spatial resolution ($\sim6\arcsec$,
$\sim18\arcsec$, and $\sim40\arcsec$ for 24, 70, and 160 \mum,
respectively) is poor compared to most optical surveys, so source
confusion is in theory a concern.  In practice, this is a sparse
enough association that source confusion among members is not a
concern.  The chances of a random alignment of a background galaxy on
top of a known association member are not large given the intense scrutiny
(including spectroscopy and deep and/or high-spatial-resolution
observations) of these association members to this point \citep[see,
e.g.,][]{lyo06}.   

\clearpage
\thispagestyle{empty}
\begin{deluxetable}{clllrclclcl}
\tabletypesize{\scriptsize}
\rotate
\tablecaption{MIPS Photometry vs. Predicted Photospheric Emission 
for the $\eta$ Cha Association \label{tab:data}}
\tablewidth{0pt}
\tablehead{
\colhead{ECHA} & \colhead{name} & \colhead{2MASS ID }& \colhead{Spectral} &
\colhead{pred.\ 24 \mum}& 
\colhead{meas.\ 24 \mum}& \colhead{pred.\ 70 \mum}& 
\colhead{meas.\ 70 \mum}& \colhead{pred.\ 160 \mum}& 
\colhead{meas.\ 160 \mum}& \colhead{L$_{IR excess}$/L$_{star}$}\\
 \colhead{number}& & & \colhead{Type\tnm{1}} & \colhead{(mJy)}&\colhead{(mJy)}& \colhead{(mJy)}
&\colhead{(mJy)}&\colhead{(mJy)}&\colhead{(mJy)}}
\startdata
01& EG Cha                    &08365623-7856454&    K6, K7.0\tnm{2}& 10.21&  10.11 $\pm$0.04& 1.2& $<$  16        &0.23& $<$ 121      &\dots\\
02& $\eta$ Cha                &08411947-7857481&    B8             & 34.54& 113.30 $\pm$0.07& 4.0&  31.1 $\pm$ 0.1&0.74& $<$ 270      & $9\times10^{-5}$ \\
03& EH Cha                    &08413703-7903304& M3.25             &  1.82&   2.23 $\pm$0.04& 0.2& $<$ 19         &0.04& $<$ 133      & $\sim10^{-6}$\\
04& EI Cha                    &08422372-7904030& M1.75             &  3.91&   5.37 $\pm$0.04& 0.5& $<$ 17         &0.09& $<$ 127      & 0.0003\\
05& EK Cha                    &08422710-7857479&    M4             &  1.27&  57.74 $\pm$0.07& 0.2&  94.8 $\pm$ 0.2&0.03& 165 $\pm$ 48 & 0.06\\
06& EL Cha                    &08423879-7854427&    M3             &  2.04&   1.86 $\pm$0.04& 0.3& $<$ 16         &0.05& $<$ 228      & \ldots \\
07& EM Cha                    &08430723-7904524&    K6, K6.9\tnm{2}&  7.77&   7.20 $\pm$0.04& 0.9& $<$  23        &0.17& $<$ 135      & \ldots \\
08& RS Cha AB                 &08431222-7904123&    A7             & 32.61&  44.86 $\pm$0.06& 3.8& $<$ 17         &0.71& $<$ 195      & \ldots \\
09& EN Cha                    &08441637-7859080&  M4.5             &  2.05&  43.80 $\pm$0.07& 0.3&  50.8 $\pm$ 0.2&0.05& $<$ 133      & 0.04 \\
10& EO Cha                    &08443188-7846311&    M1             &  3.40&   2.81 $\pm$0.04& 0.4& $<$ 17         &0.08& $<$ 180      & \ldots \\
11& EP Cha                    &08470165-7859345&  K5.5, K6.5\tnm{2}&  7.14& 198.20 $\pm$0.09& 0.8& 184.7 $\pm$ 0.2&0.16& 208 $\pm$ 48 & 0.04\\
12& EQ Cha                    &08475676-7854532& M3.25             &  4.59&   4.33 $\pm$0.05& 0.6& $<$ 19         &0.11& $<$ 122      & \ldots\\
13& HD 75505                  &08414471-7902531&    A1             & 11.92&  10.41 $\pm$0.04& 1.4& $<$ 18         &0.26& $<$ 137      & \ldots \\
14& ES Cha, ECHA J0841.5-7853 &08413030-7853064& M4.75             &  0.45&   8.63 $\pm$0.04& 0.1& $<$ 17         &0.01& $<$ 173      & 0.04 \\
15& ET Cha, ECHA J0843.3-7905 &08431857-7905181& M3.25             &  1.79& 232.50 $\pm$0.11& 0.2& 173.3 $\pm$ 0.3&0.04& $<$ 121      & 0.19\\
16& ECHA J0844.2-7833         &08440914-7833457& M5.75             &  0.25&   8.68 $\pm$0.08&0.03&  \ldots        &0.01&  \ldots      & 0.04\\
\enddata
\tnt{1}{Spectral types from \citet{ls04} except as explained in note 2}
\tnt{2}{Spectral types from \citet{lyo04} used for improved model fits as explained in section \ref{irxs}}
\end{deluxetable}

\clearpage

\begin{figure}
\epsscale{1.0}
\plotone{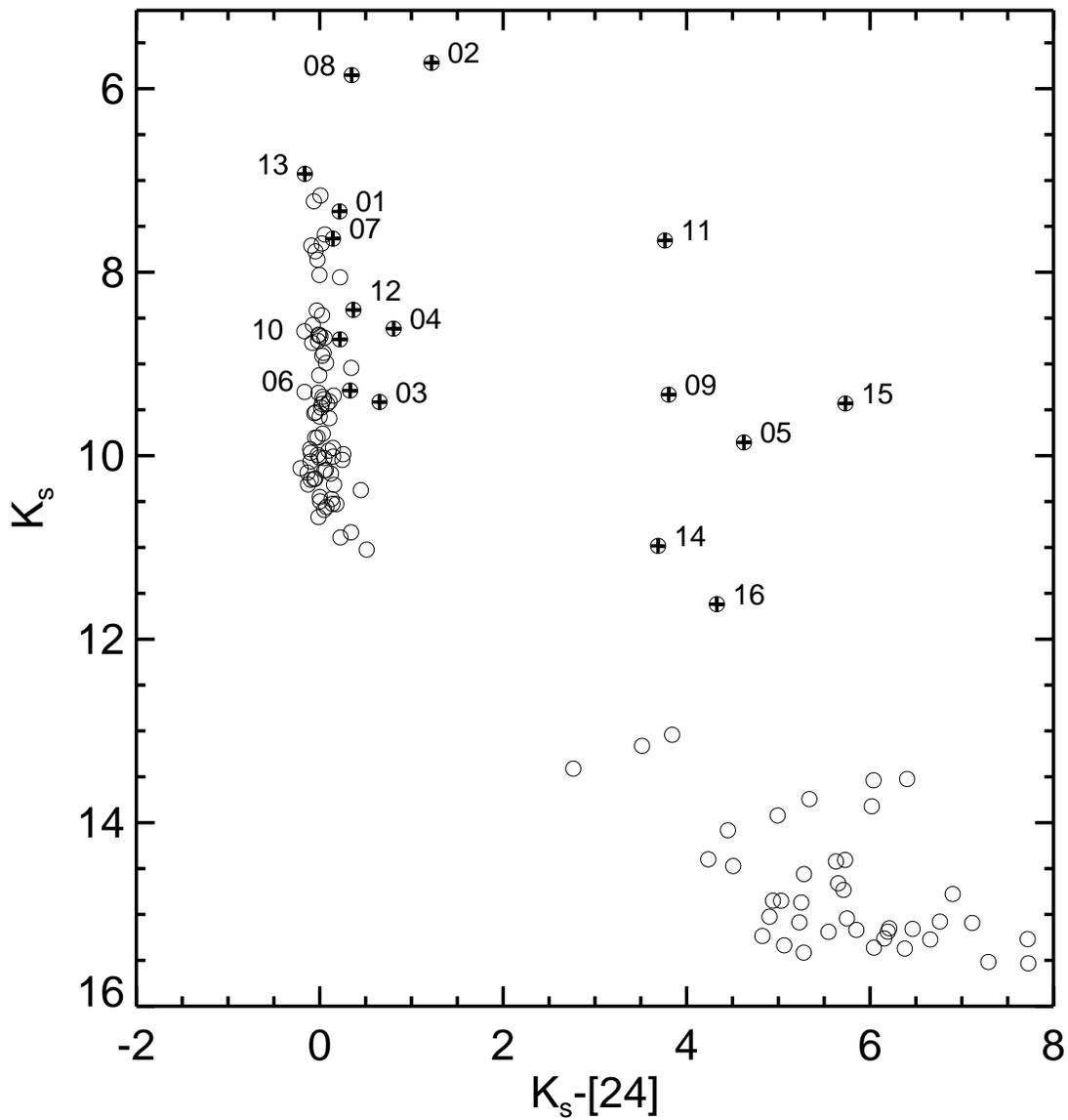}
\caption{
\ks\ magnitude vs.\ \ks$-$[24] color for all 153 MIPS-24 sources detected
in our map with 2MASS \ks\ counterparts.  The known members are indicated
by a '+' sign within the circular symbol.  Extragalactic objects are clustered 
in the clump at the lower right and stars without 24 \mum\ excesses lie in the 
vertical grouping at \ks$-[24]\sim$0 from \ks$\sim$6-11. 
\label{kk24}}
\end{figure}

\clearpage

\section{Results}

We analyzed the \ks$-$[24] and \ks$-$[70] colors of our sources and compared 
the observed spectral energy distributions (SEDs) of the known association members 
with model photospheric spectra to look for signs of infrared excess.

\subsection{\ks\ vs.\ \ks$-$[24]}
\label{kkm24sec}
 
Fig.~\ref{kk24} shows the \ks\ vs.\ \ks$-$[24] diagram for the 129 
objects we detected at 24 \mum\ with \ks\ counterparts in 2MASS. 
In this figure most stellar photospheres are near
\ks$-$[24]$\sim$0 and the clump of objects at the lower right are
likely galaxies.  Six of the known $\eta$ Cha members fall in
the upper right of this diagram, showing strong 24 \mum\ excess:
ECHA 05, 09, 11, 14, 15, and 16.  $\eta$ Cha itself (ECHA 02) has 
a clear excess.  Nine other members fall in or just redward of
the ``photospheric'' strip in the \ks\ vs.\ \ks$-$[24] diagram.
This red trend is not due to observational errors. We examined the scatter 
in \ks$-$[24] for stars with good measured S/N at 24 \mum\
corresponding to \ks$\lesssim$10.  Among the 46 stars that are not 
known members, \ks$<$10, and \ks$-[24]<$0.5, the mean \ks$-$[24] 
color is 0.016$\pm$0.098 mag.  We take this to indicate that 
our measurement error in the color is $\sim$0.1 mag.  Therefore, 
the majority of known members just redward of the photospheric 
strip have clearly non-zero \ks$-$[24] colors.  

\clearpage
\begin{figure}
\epsscale{1.0}
\plotone{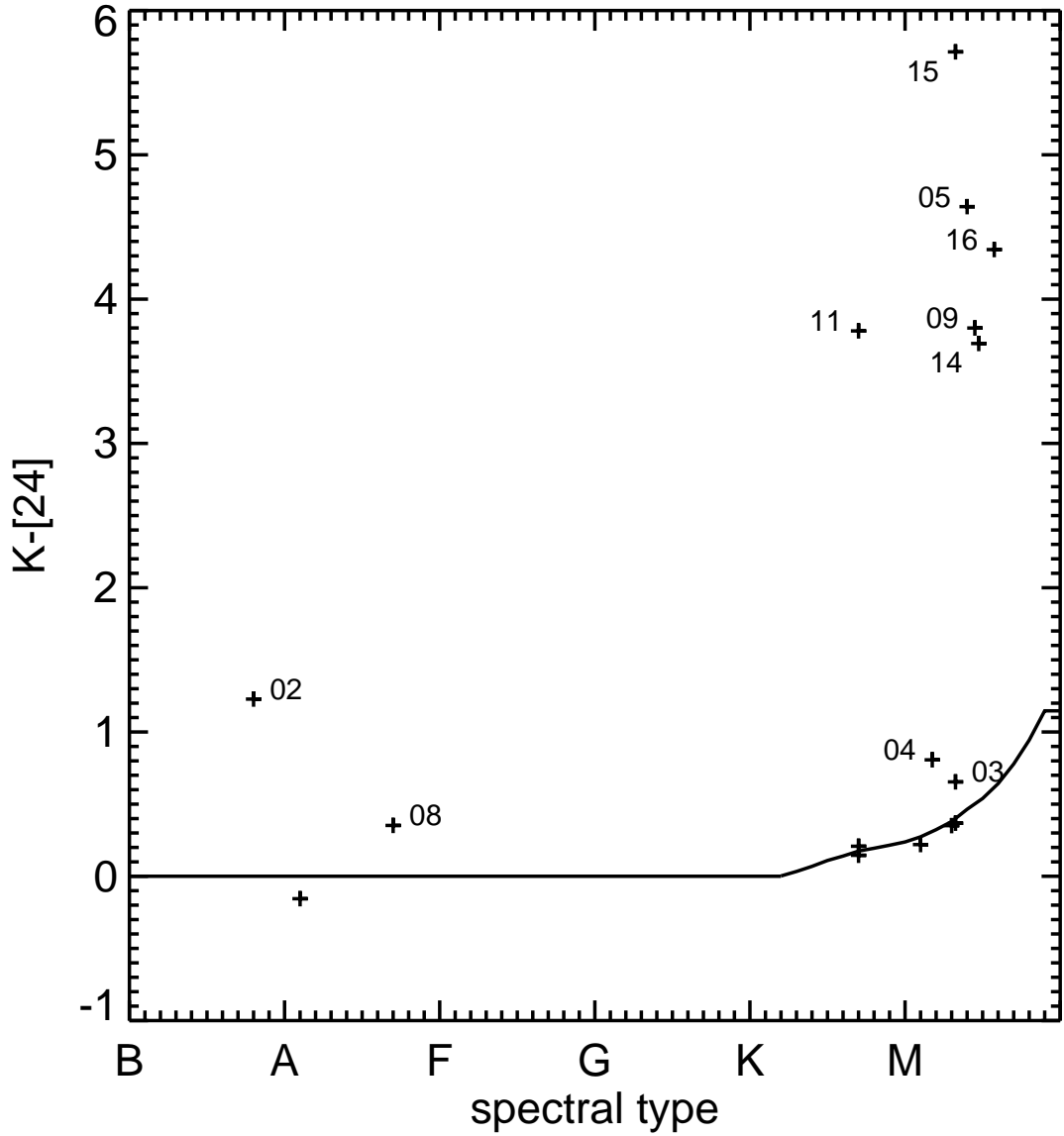}
\caption{
\ks$-$[24] vs.\ spectral type for all  $\eta$ Cha association members; 
members that are 24 $\mu$m excess candidates are labeled by their ECHA number.
The solid line represents the locus of stellar photospheric colors
determined from a study of nearby stars by \citep{gaut07}.
\label{spty}}
\end{figure}
\clearpage

To understand whether these nine association members near the 
photospheric strip have infrared excess 
it is necessary to take into account the spectral types of the 
individual objects.  While the \ks$-$[24] color is near zero 
for most stellar photospheres, it becomes non-zero and varies
with spectral type among M stars \citep{gaut07}.  Since many of our
objects are of spectral type M this effect is important.
Figure~\ref{spty} plots \ks$-$[24] vs.\ spectral type for all 
association members.  The locus of luminosity class V photospheres 
was obtained from \citet{lang91} for types A to K5 and from \citet{gaut07} 
(and references therein) for K6 through M9. The strong excess sources 
are still obvious at the top right of the diagram.  

Six of the nine association members fall on the locus of photospheric 
\ks$-$[24] colors and thus do not possess infrared excess.  However, 
three objects (ECHA 03, 04, and 08) show a clear infrared excess.  
Their \ks$-$[24] uncertainty combined with the dispersion of normal dwarf 
stars about the plotted locus places ECHA 03 about $5\sigma$ away and 
ECHA 04 and 08 about $8\sigma$ away from the photospheric locus. Their observed 
24 \mum\ flux densities are 1.23, 1.37, and 1.38 times greater than 
expected for their spectral type based on our model spectra 
(see Table~\ref{tab:data}). 

In our large $0.5\arcdeg\times0.5\arcdeg$ survey field, the possibility 
exists that previously unrecognized members of the $\eta$ Cha association
might be revealed as stars with \ks$-$[24] excess.  Figure~\ref{kk24} 
shows that no obvious new candidates are detected.  While objects in
the clump of likely galaxies have \ks$-$[24] colors comparable to other
$\eta$ Cha members with excess, all are fainter than \ks$\sim$13 and thus 
unlikely to be stars at a distance of 100 pc.  This result is consistent 
with recent deep imaging and spectroscopic studies of the region that have
also failed to identify additional members \citep[see, e.g.,][]{lyo06}.

\clearpage
\begin{figure}
\epsscale{1.0}
\plotone{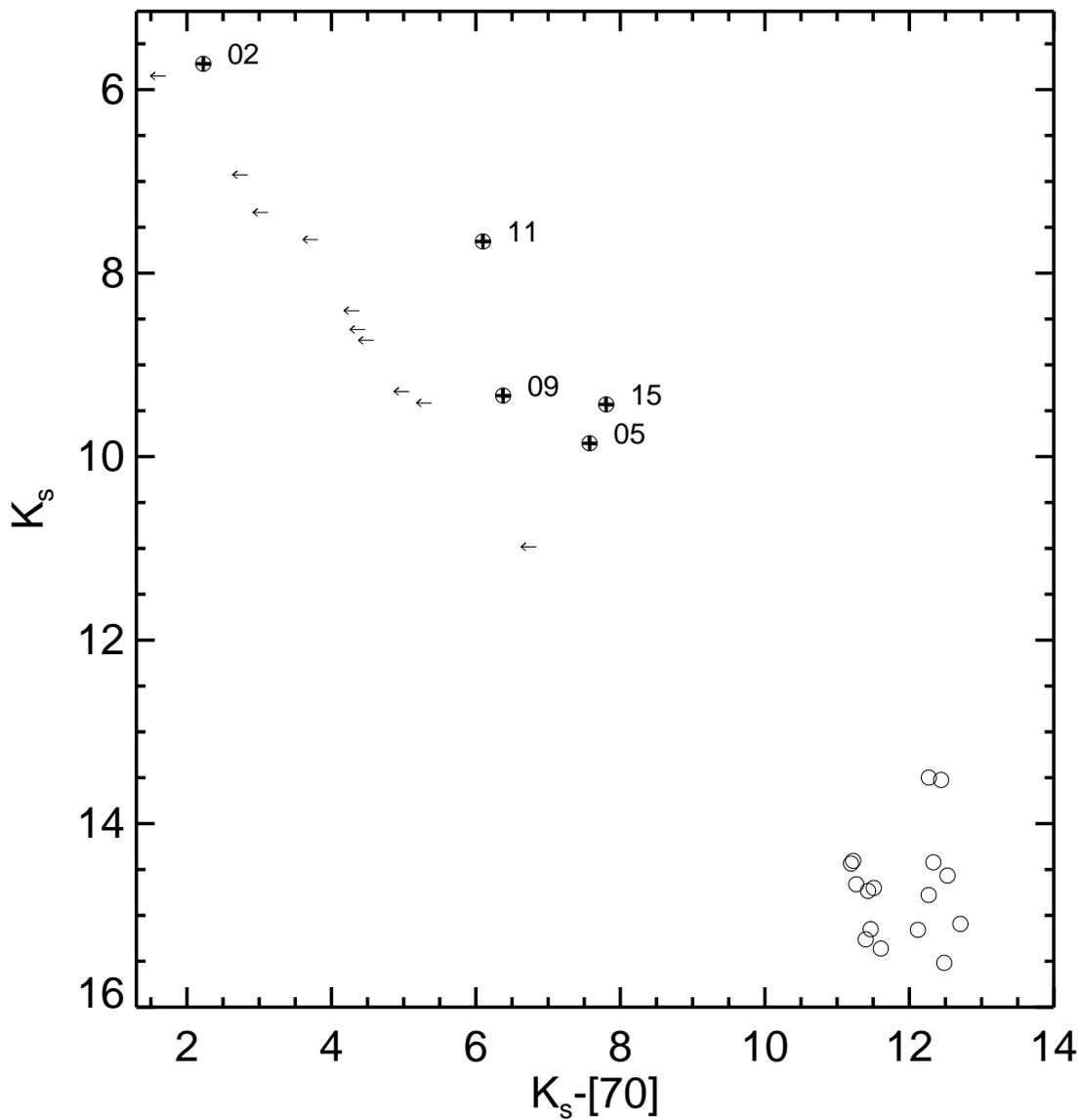}
\caption{ 
\ks\ magnitude vs. \ks$-$[70] color for all 21 MIPS-70 sources with
2MASS \ks\ counterparts.  The known members have an additional embedded
+ sign within the circular symbol. Extragalactic objects are clustered in
the clump at the lower right and stars without 70 \mum\ excesses would
lie near \ks$-[70]\sim$0, were any detected.  Most of the association
members are undetected at 70 microns, with upper limits to their
\ks$-$[70] color indicated by leftward pointing arrows.
\label{kk70}}
\end{figure}
\clearpage

\subsection{\ks\ vs.\ \ks$-$[70] and 160 \mum}

Figure~\ref{kk70} shows the \ks\ vs.\ \ks$-$[70] diagram for the 21 
objects detected at 70 \mum\ that have \ks\ counterparts in 2MASS. 
As in Figure~\ref{kk24}, the clump of objects at the lower
right are likely galaxies.  Our observations were not sensitive 
enough to detect stellar photospheres at the distance of $\eta$ Cha, 
and thus most of the members are not detected (upper limits are shown
as arrows).  However, five stars show strong 70 \mum\ excess: 
ECHA 2, 5, 9, 11, and 15.  All five also have 24 \mum\ excess.
Our 160 \mum\ observations were also not sensitive enough to detect 
stellar photospheres so the two stars detected, ECHA 5 and 11, show 160 
\mum\ excess.

\clearpage
\begin{figure}
\epsscale{0.9}
\plotone{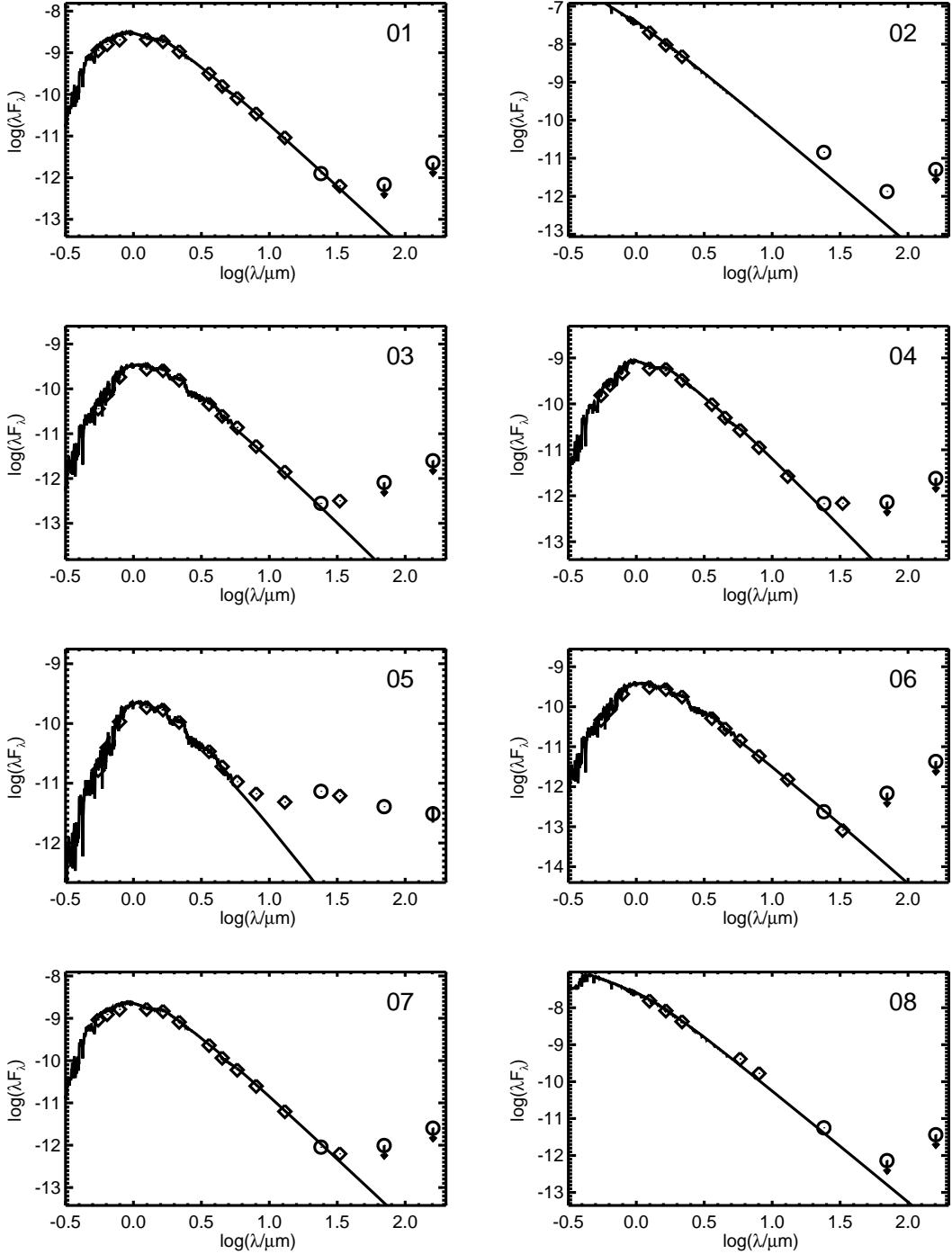}
\caption{
SEDs of the known members of the $\eta$ 
Cha association.  The $x$-axis is log of the wavelength in microns; the 
$y$-axis is log of $\lambda F_{\lambda}$ in cgs units.  Diamonds are points from 
the literature; circles are MIPS points where upper limits are indicated by
downward-pointing arrows.  For ECHA 11, additional IRAS PSC points are
given (grey squares) for comparison.  For each source, a model 
representing the stellar photospheric emission is shown as a solid
line.
\label{sed_fig1}}
\end{figure}

\begin{figure}
\epsscale{0.9}
\plotone{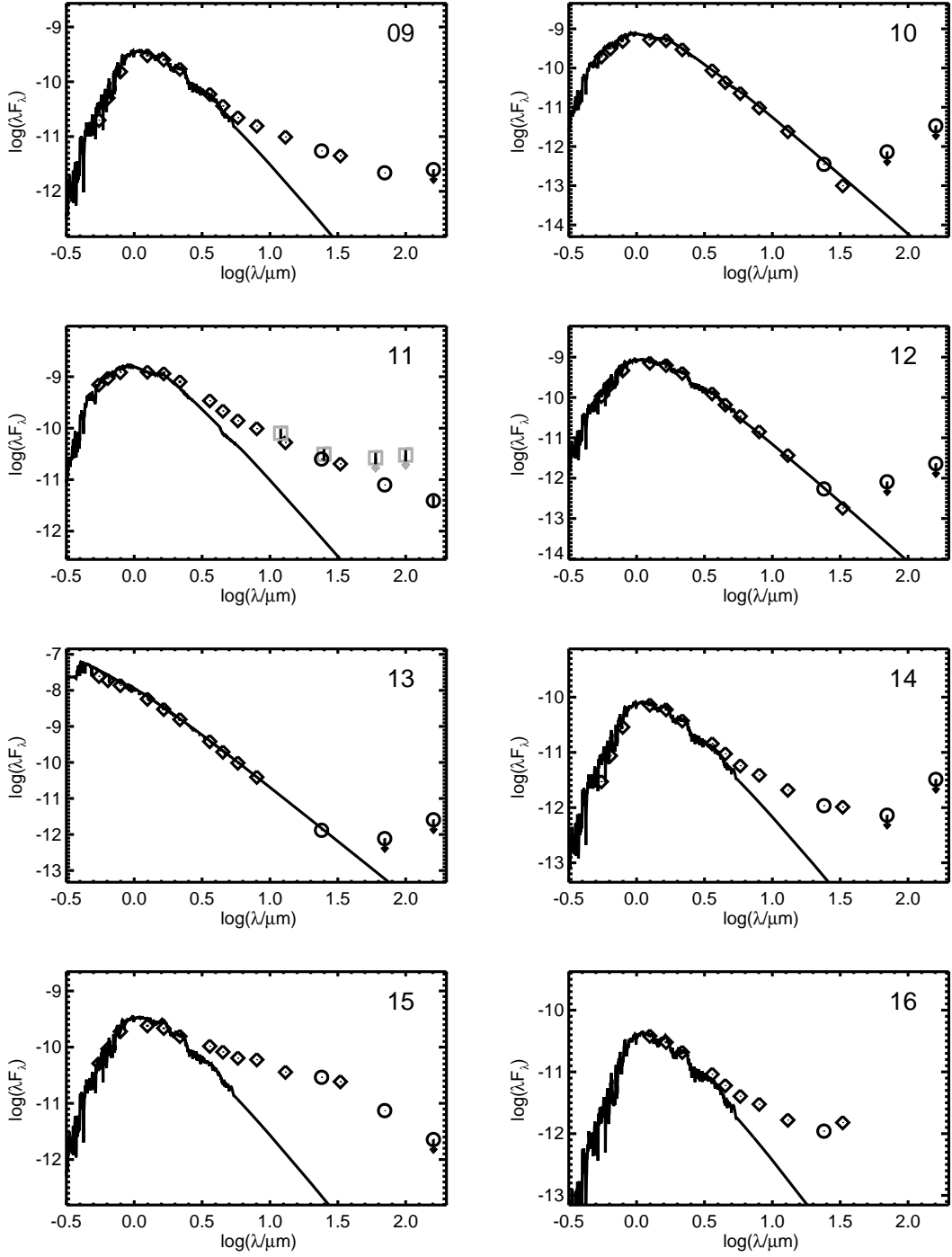}
\caption{SEDs of the known members of the $\eta$ Cha association, continued.
\label{sed_fig2}}
\end{figure}
\clearpage

\subsection{Infrared Excesses of $\eta$ Cha Members}
\label{irxs}

Figures~\ref{sed_fig1} and \ref{sed_fig2} present the spectral energy
distributions (SEDs) for all of the known members of the $\eta$ Cha 
Association covered by our MIPS map. Spectral types from Table \ref{tab:data} 
were used to select the model spectra as described in section \ref{obs}. 
The spectral types from \citet{lyo04} were 
used for the K type stars, ECHA 1, 7 and 11, as models of the \citet{ls04} 
types are clearly too blue for these stars and would have required 
reddening from extinction much greater than allowed by the results of 
\citet{lyo04} and \citet{ls04}.

Normalization of the model spectrum at \ks\ for ECHA 11 did not produce a 
consistent fit to the measurements at wavelengths shorter than \ks. We 
therefore normalized ECHA 11 at J, revealing an excess that begins at \ks.

In one object, ECHA 08 (RS Cha), an apparent infrared excess is seen at 
5.8, 8.0, and 24 \mum.  This source is a known eclipsing binary with
a peak-to-peak amplitude of 0.75 mag \citep{claus80}.  The ephemeris 
in Clausen \& Nordstrom indicates that the MIPS and IRAC observations 
were both made outside of eclipse at times when the b magnitude of RS Cha 
differed by about 0.03. The 2MASS observation was made in the 
secondary eclipse when RS Cha's b magnitude was about 0.2 fainter than 
for the MIPS observation. Given the zero [5.8]-[8.0] color found in 
simultaneous observations of this source by IRAC \citep{meg05}, and the 
0.1 mag slight blue [8.0]-[24] color indicated by our data -- both 
consistent with a stellar photosphere -- it appears that our apparent 
\ks$-$[24] excess of 0.35 mag is an artifact of non-simultaneous photometry.  

Another object, ECHA 13, the A1 star HD 75505, is reported by \citet{ls04} to 
be slightly reddened relative to the other ECHA objects. \citet{lyo04} also 
finds HD 75505 slightly reddened and, based on this reddening and a small 
K$-$L excess reported by \citet{lyo03}, attributes the reddening to an 
edge-on circumstellar disk. We find no evidence for a \ks$-$[24] excess in 
HD 75505 that might be expected from such a disk. While the absence of a 
24 \mum\ excess is not evidence of no circumstellar disk, disk material 
capable of producing the reddening would have to be at a large distance 
from an A star to be invisible at 24 \mum. We note that \citet{meg05} did 
not find a significant \ks$-$[3.6], a color similar to K$-$L, excess for HD 75505. 

The availability of the MIPS data now makes it possible to calculate
the fractional infrared excess luminosity for each member of the $\eta$ 
Cha association.  Optical and near-infrared points were used to define 
the stellar photospheres, which were represented as a blackbody at the 
known effective temperature of each source.  Integration of the infrared 
excess across the Spitzer bands was done done using cubic spline interpolation 
to the observed mid-infrared and far-infrared data points, and subtracting 
off the photospheric contribution point-by-point.  To account for the excess
luminosity out to submillimeter wavelengths, a blackbody extrapolation was
performed from the longest available infrared data point, with a blackbody
temperature chosen using Wien's law for that wavelength.  The results of 
this analysis appear in the far right column of Table~\ref{tab:data}.

\section{Disk Properties in the $\eta$ Cha Association}

The infrared excesses reported in the literature at some wavelength between 
6-160 \mum\ in nine association members require that circumstellar dust be 
present within 10 AU of these stars.  A flattened disk is the only dust configuration
that could be dynamically stable at the age of the association.  From
this point forward, we assume that the infrared excesses in the $\eta$ Cha
association are produced by dusty circumstellar disks.  A tally of our MIPS 
results shows 9 of 16 association members have excesses at 24 \mum\ and at least 
5 of the members with 24 \mum\ excess also have a 70 \mum\ excess. We compare our results 
to disk indicators from other observations in $\eta$ Cha.

\subsection{Spitzer/IRAC}

\citet{meg05} found excesses between 3.6-8.0 \mum\ in six stars: 
ECHA 5, 9, 11, 14, 15, and 16.  All of these also have excesses in
our data at $\lambda\ge$24 \mum.  There are two stars, ECHA 3 and 4, 
where we have found MIPS excess but no mid-infrared excess is detected.  
This is not surprising, as these stars have very subtle MIPS-24 excesses.
The steep brightening of the stellar photosphere toward shorter
wavelengths would rapidly obscure such a modest mid-infrared excess of 
the strength seen at 24 \mum.  No IRAC photometry was obtained for 
\citet{meg05} for ECHA 2, so the inner disk of this source is uncharacterized.
The frequency of IRAC excess in the association is 6/17 $=$ 35\%. 
We note for completeness that \citet{meg05} discussed in detail a 
comparison between IRAC photometry and the L-band photometry from 
\citet{lyo03}; so we do not repeat it here.

\subsection{Spitzer/IRS}
\label{spitirs}

\citet{bou06} report 13 and 33 \mum\ flux densities for fifteen 
$\eta$ Cha members, based on Spitzer low-resolution spectra. The 
same six objects showing IRAC excess above also show 13 \mum\ excess. 
In addition, they found 33 \mum\ excess in ECHA 3 and 4 Ð-- two stars 
for which we have found weak 24 \mum\ excess. For ECHA 7, \citet{bou06} 
report a 33 \mum\ flux density of 6.9 mJy, and a 33 \mum\ to 13 \mum\ 
flux ratio noticeable larger than for their other diskless stars, but 
do not classify this as an excess. Our photosphere model is in excellent 
agreement with the flux densities measured with IRS at 13 \mum\ and 
with MIPS at 24 \mum. It predicts a flux density of 4.1 mJy at 33 \mum. 
The 40\% difference between our prediction and the reported 33 \mum\ 
flux is significantly larger than the reported errors in the IRS 
measurements and argues that ECHA 7 does, in fact, show a 33 \mum\ 
excess. In view of this disagreement on interpretation of 33 \mum\ 
flux of ECHA 7 we examined the now public Spitzer data on which 
Bouwman et al. based their analysis. We find that the spectrum of 
the ECHA 7 spectrum beyond 20 \mum\ appears to be shallower than 
the Rayleigh-Jeans spectrum expected for photospheres in this region 
of the spectrum. A ratio of the ECHA 7 spectrum to the ECHA 1 spectrum, 
a star with a low value of 33 to 13 \mum\ flux ratio from the same 
IRS cluster observation, also shows that ECHA 7 has significantly 
shallower slope than ECHA 1 beyond 20 \mum. Given this evidence, we 
conclude that ECHA 7 probably has a weak 33 \mum\ excess indicative of a 
disk with a large inner hole. The MIPS 70 \mum\ upper limit is consistent with 
this interpretation. The overall frequency of 13 \mum\ excess in 
$\eta$ Cha is 6/15 $=$ 40\%; the revised frequency of 33 \mum\ excess, 
including ECHA 7, is 9/15 $=$ 60\%.
 
\subsection{Disk Structure}

Table~\ref{tab:disks} summarizes the results for disks
in the $\eta$ Cha association.  Ten members show infrared excesses 
attributable to disks at one or more wavelengths between 2.2 and 
160 \mum; nine have excesses detectable by MIPS at 24 \mum\ and one 
has an excess beginning at 33 \mum.  In ECHA 11, 15, and 16, excess
is detected from 3.5-33 \mum\ (with 11's excess extending even to 
2.2 \mum), indicating that these objects possess
continuous disks extending from near the stellar surface to radii
$\gtrsim$10s of AU.  Remarkably, six of the other disks show evidence 
for inner holes.  ECHA 9 and 14 have small inner holes diagnosed by
the presence of excess only for $\lambda \ge$ 6 \mum; ECHA 5 is similar,
with the excess starting at 8 \mum.  The SEDs of ECHA 3 and 4 require
larger disk inner holes, with excess absent at 13 \mum\ but detected
at 24 \mum.  Finally, the disk of ECHA 7 has the largest inner hole 
of all, with a 33 \mum\ excess but no 24 \mum\ excess.  The tenth
star with a disk is $\eta$ Cha itself; while the available data does
not allow the presence of an inner hole to be discerned, the large ratio
of the 24 and 70 \mum\ excess fluxes ($\sim$ 3; Table~\ref{tab:data})
is characteristic of warm material near 250 K.

\subsection{Disks and Mass Accretion}

\cite{law04} studied the H$\alpha$ emission in ten members of the
association, and found that four show high-velocity line wings
indicative  of gas accretion: ECHA 5, 9, 11, and 15.  \citet{meg05} 
reanalyzed H$\alpha$ data from Song  \etal\ (2004) after the method of
White \& Basri (2003), and found that ECHA 16 also shows signs of gas
accretion.  We find that all five of these have circumstellar disks
with L$_d$/L$_* \ge$ 0.04. The two disks around ECHA 3 and 4, where 
Spitzer has found weak 24 and 33 \mum\ excesses, showed no accretion 
signatures. These results support the long-known correlation between 
the strength of disk excess and accretion rates.  H$\alpha$ studies of 
ECHA 2 and 14 are still needed to see if this relation holds for all 
the disks Spitzer has detected in $\eta$ Cha.

\clearpage
\begin{deluxetable}{ccccc}
\tabletypesize{\scriptsize}
\tablecaption{Disk indicators in $\eta$ Cha Members. \label{tab:disks}}
\tablewidth{0pt}
\tablehead{
\colhead{ECHA} & \colhead{H$\alpha$ \mdot\tablenotemark{a}} & 
\colhead{\citet{meg05}}& \colhead{\citet{bou06}} &
\colhead{this work}\\
\colhead{(=LS)} & \colhead{} & \colhead{IRAC} &
\colhead{IRS} & \colhead{MIPS} }
\startdata
ECHA 01& (no data)            & no        & no         & no  \\ 
ECHA 02& (no data)            & (no data) & (no data) & yes \\ 
ECHA 03& no                   & no        & yes       & yes \\ 
ECHA 04& no                   & no        & yes       & yes \\ 
ECHA 05& yes                  & yes       & yes       & yes \\ 
ECHA 06& no                   & no        & no        & no  \\ 
ECHA 07& no                   & no        & yes\tablenotemark{b}& no  \\ 
ECHA 08& (no data)            & no        & (no data) & no\tablenotemark{c}  \\ 
ECHA 09& yes                  & yes       & yes       & yes \\ 
ECHA 10& no                   & no        & no        & no  \\ 
ECHA 11& yes                  & yes       & yes       & yes \\ 
ECHA 12& no                   & no        & no        & no  \\ 
ECHA 13& (no data)            & no        & (no data) & no  \\ 
ECHA 14& (no data)            & yes       & yes       & yes \\ 
ECHA 15& yes                  & yes       & yes       & yes \\ 
ECHA 16& yes                  & yes       & yes       & yes \\ 
ECHA 17& (no data)            & no        & no        & (no data) \\ 
ECHA 18& (no data)            & no        & no        & (no data) \\ 
\enddata 
\tablenotetext{a}{All analysis in this column from \citet{law04},
except ECHA 16, for which Megeath \etal\ (2005) reanalyzed data from Song
\etal\ (2004) after the method of White \& Basri (2003).}
\tablenotetext{b}{This excess was not claimed in \citet{bou06}. See 
section \ref{spitirs} for explanation.}
\tablenotetext{c}{Apparent excess not attributable to a disk; see
text in section \ref{irxs}.}

\end{deluxetable}
\clearpage

\subsection{Disks and Binarity}

Seven stars in the $\eta$ Cha association are known multiples: ECHA 1,
7, 8, 9, 12, 17, and 18 \citep[][and references therein]{bran06,bou06}.  
Infrared excess is  detected for ECHA 7 and 9;
in ECHA 9, the disk must be circumbinary. Of the eleven single stars,
excess is detected around six.  While an  anticorrelation between
disks and binarity is suggested by these results,  small number
statistics prevent any meaningful conclusion from being drawn.

\section{$\eta$ Cha and Other Young Stellar Groups}

Two other well-studied stellar associations close in age to $\eta$ Cha
also have MIPS studies: the TW Hydra Association \citep[TWA;][]{low05}, 
and the Beta Pic Moving Group \citep[BPMG;][]{bpmgp}.  At 8-10 Myr, TWA 
is thought to be comparable in age to $\eta$ Cha, and BMPG is slightly 
older at $\sim$12 Myr.  \citet{low05} find for TWA that there are very 
large excesses around four of the TWA stars, with possibly a subtle 24 
\mum\ excess around one more of the stars.  \citet{bpmgp} re-reduced the 
TWA MIPS data in exactly the same fashion as here in $\eta$ Cha and in the 
BPMG, and find the TWA disk fraction at 24 \mum\ to be 7/23 stars, or 30\%.
\citet{bpmgp} find that the BPMG has a 24 \mum\ disk fraction of 7 of
30 stars, or 23\%.  The larger disk fraction we find at 24 \mum\ for
$\eta$ Cha, 56\%, suggests that it may actually be younger than either 
of these other two associations. Since so many association members are 
undetected at 70 \mum, the best constraint we can put on the disk fraction 
at this wavelength is a lower limit of 31\%. We note that the various 
observational studies of the $\eta$ Cha association are not unbiased 
with respect to the known association members. \citet{bou06} excludes 
the early type members and this paper leaves two objects at 24 \mum\ 
and three objects at 70 \mum\ unobserved, with many 70 \mum\ upper limits 
for those that were observed. However this should not affect 
conclusions based on a larger disk fraction in $\eta$ Cha than in the 
other associations.

Within our $\eta$ Cha association members, there are three groups of
24 \mum\ excess sources attributable to disks.  There are stars without 
excesses at 24 \mum\ (ECHA 1, 6, 7, 8, 19, 12 and 13), stars with small 
excesses (ECHA 2, 3 and 4), and
stars with relatively large excesses (ECHA 5, 9, 11, 14, 15 and 16). This
is reminiscent of the distribution of TWA 24 \mum\ excesses found by
\citet{low05}, where there are either large or subtle 24 \mum\
excesses.  In our results from $\eta$ Cha, the largest \ks$-$[24]
values that we find are those for ECHA 15 (5.73), ECHA 05 (4.62) and
ECHA 16 (4.33). ECHA 9, 11, and 14 all have \ks$-$[24]$\sim$3.8, and
the rest are all $<$1.2. These largest values are close to, but still
below, the four extreme TWA stars (\ks$-$[24]=5.8, 5.0, 4.4, and 4.4 for
TWA 1, 3, 4, and 11, respectively) from \citet{low05}.  In contrast,
\citet{bpmgp} find that the BPMG has no extreme excesses -- the
largest \ks$-$[24] they find is for $\beta$ Pic itself at only 3.5.
We can compare the ratios of measured to predicted fluxes at 24 \mum\
for each of these three clusters.  The median $F_{\rm meas}/F_{\rm
pred}$ for all MIPS disks is 51.4, 21.2, and 1.83 for TWA, ECHA, and
the BPMG, respectively.

We conclude that, while in terms of 24 \mum\ disk fraction, the $\eta$ Cha
association is younger than both TWA and the BPMG, it is solidly
intermediate in disk properties between the two clusters.

Fractional infrared luminosities are shown for nine $\eta$ Cha members
in Table~\ref{tab:data}.  Three sources ($\eta$ Cha itself, ECHA 3 and 
ECHA 4) have fractional infrared luminosities $\le 0.001$, which
are typical for debris disks around main sequence stars (Bryden \etal\
2006; Su \etal\ 2006).  The classical T Tauri star ECHA J0843.3-7905 
(ECHA 15) has an L$_d$/L$_*$ of nearly 20\%, consistent with an
optically  thick YSO disk like that of TW Hya.  Five other members 
(ECHA 5, 9, 11, 14, and 16) have fractional infrared
luminosities of  0.04-0.06.  These disks are particularly interesting,
as their luminosities  are intermediate between those of optically
thick young disks and those  of debris disks.  Other examples of disks
with L$_d$/L$_*$ in this range  have recently been found in Spitzer
studies of weak-line T Tauri stars  (Padgett \etal\ 2006; Cieza \etal\
2007).  Such disks can be explained by  one of two models: an
optically thick disk that has been highly flattened  by settling of
dust to its midplane, or an optically thin disk that is 
$\sim$10$\times$ denser than the prototypical young disk of $\beta$
Pictoris.   Sensitive submillimeter observations will be needed to
distinguish between  these two possibilities.  In either case, these
five disk systems systems  represent a key transitional stage in disk
evolution; further detailed  studies are needed.

\section{Conclusions}

Our Spitzer/MIPS data, combined with a reanalysis of IRAC and
IRS data, indicates an overall mid-infrared disk frequency of at least 10/18 $=$
56\% in the $\eta$ Cha association at ages of $\sim$ 8 Myrs.  This is
significantly higher than the disk fraction observed in the TW Hydra
or $\beta$ Pictoris associations at similar ages.

The $\eta$ Cha disks show clear transitional characteristics
between young stellar object and debris disks, both in terms of the
distribution of their fractional infrared luminosities and the
presence of inner holes (diagnosed by the spectral energy
distribution) in  6 of the 10 disks studied.

\acknowledgements

This work is based on observations made with the Spitzer Space 
Telescope, which is operated by the Jet Propulsion Laboratory,
California  Institute of Technology under contract with NASA. Support
for this work  was provided by NASA through contract 1255094.  
Gautier was also partially supported under the Research and Technical 
Development funds at the Jet Propulsion Laboratory. This
work also makes use of  data products from the Two Micron All Sky
Survey, which is a joint project  of the University of Massachusetts
and the Infrared Processing and Analysis  Center/California Institute
of Technology, funded by the National Aeronautics  and Space
Administration and the National Science Foundation.


\begin{thebibliography}

\bibitem[Bouwman \etal(2006)]{bou06} Bouwman, J., \etal, 2006, \apjl, 653, L57
\bibitem[Brandeker \etal(2006)]{bran06} Brandeker, A., Jayawardhana, R., 
Khavari, P., Haisch, K.E., Mardones, D. 2006, Ap.J., 652, 1572
\bibitem[Bryden \etal(2006)]{bryden06} Bryden, G., \etal, 2006, \apj, 646, 1038
\bibitem[Cieza \etal(2007)]{cieza07} Cieza, L., \etal, 2007, arXiv/0706.0563
\bibitem[Clausen \& Nordstrom (1980)]{claus80} Clausen, J.V. and Nordstrom, 
B. 1980, \aap, 83, 339
\bibitem[Engelbracht \etal(2007)]{chad07} Engelbracht, C., \etal, 2007, \pasp, in press
\bibitem[Gautier \etal(2007)]{gaut07} Gautier, T. N., \etal, 2007, \apj, 667, 527 
\bibitem[Gordon \etal(2005)]{gor05} Gordon, K., \etal, 2005, \pasp, 117, 503
\bibitem[Gordon \etal(2007)]{gor07} Gordon, K., \etal, 2007, \pasp, in press
\bibitem[Haisch \etal(2005)]{hach05} Haisch, K.E. Jr., Jayawardhana, R., and Alves, J. 2005 \apj, 627 L57
\bibitem[Lang(1991)]{lang91} Lang, K. R. 1991, Astrophysical Data, Springer-Verlag, New York
\bibitem[Lawson \etal(2002)]{law02} Lawson, W.A., Crause, L.A., Mamajek, E.E., and Feigelson, E.D. 2002 \mnras, 329 L29
\bibitem[Lawson \etal(2004)]{law04} Lawson, W.A., Lyo, A-R. and Muzerolle, J. 2002 \mnras, 351 L39
\bibitem[Lejeune \etal(1997)]{lej97} Lejeune, T., Cuisinier F., \& Buser R. 1997, \aaps, 125, 229
\bibitem[Low \etal(2005)]{low05} Low, F., \etal, 2005, \apj, 631, 1170
\bibitem[Luhman \& Steeghs(2004)]{ls04} Luhman, K.L. and Steeghs, D. 2004, \apj, 609, 917
\bibitem[Lyo \etal(2003)]{lyo03} Lyo, A-R., \etal, 2003, \mnras, 338, 616
\bibitem[Lyo \etal(2004)]{lyo04} Lyo, A-R., \etal, 2004, \mnras, 355, 363
\bibitem[Lyo \etal(2006)]{lyo06} Lyo, A-R., \etal, 2006, \mnras, 368, 1451
\bibitem[Makovoz \& Marleau(2005)]{mak05} Makovoz, D., \& Marleau, F., 2005, \pasp, 117, 1113
\bibitem[Mamajek \etal(1999)]{mama99} Mamajek, E.E., Lawson, W.A., and Feigelson, E.D. 1999, \apj, 516 L77
\bibitem[Mamajek \etal(2000)]{mama00} Mamajek, E.E., Lawson, W.A., Feigelson, E.D. 2000, \apj, 544, 356
\bibitem[Megeath \etal(2005)]{meg05} Megeath, S.T. \etal, 2005 \apj, 634 L113
\bibitem[Moshir \etal(1992)]{moshir}Moshir, M., Kopman, G., Conrow, T. 1992, 
IRAS Faint Source Survey and Explanatory Supplement
\bibitem[Padgett \etal(2006)]{padgett06} Padgett, D.L., \etal, 2006, \apj, 645, 1283
\bibitem[Rieke \etal (2004)]{rieke04} Rieke, G., \etal, 2004, \apjs, 154, 25
\bibitem[Rebull \etal(2007a)]{perp} Rebull, L., \etal, 2007a, \apjs, 171, 447
\bibitem[Rebull \etal(2007b)]{bpmgp} Rebull, L., \etal, 2007b, \apj, submitted
\bibitem[Skrutskie \etal(2006)]{2mass} Skrutskie, M., \etal, 2006, \aj, 131, 1163
\bibitem[Song \etal (2004)]{song04} Song, I., \etal, 2004, ApJ, 600, 1016. 
\bibitem[Su \etal (2006)]{su06} Su, K.Y.L. \etal, 2006, \apj, 653, 675
\bibitem[Werner \etal(2004)]{wer04} Werner, M., \etal, 2004, \apjs, 154, 1
\bibitem[White \& Basri(2003)]{wb03} White, R., \& Basri, G., 2003, \apj, 582, 1109
\end{thebibliography}
\end{document}